\documentclass[prl,floatfix,twocolumn,showpacs]{revtex4}
\usepackage{amsmath}
\usepackage{amsfonts}
\usepackage{mathrsfs}
\usepackage{epsfig}
\usepackage{graphicx}
\usepackage{dcolumn}

\begin{document}
\title{Extreme variability in convergence to structural balance in
frustrated dynamical systems}
\author{Rajeev Singh$^1$, Subinay Dasgupta$^2$ and Sitabhra Sinha$^1$}
\affiliation{$^1$The Institute of Mathematical Sciences, CIT Campus,
Taramani,
Chennai 600113, India.\\
$^2$Department of Physics, University of Calcutta, 92 Acharya Prafulla
Chandra Road, Kolkata 700009, India.}
\date{\today}
\begin{abstract}
In many complex systems, the dynamical evolution of the different
components can result in adaptation of the connections between them.
We consider the problem of how a fully connected network of
discrete-state dynamical elements 
which can interact via positive or negative links, approaches
structural balance by evolving its links to be consistent with the
states of its components. 
The adaptation process, inspired by Hebb's
principle, involves the 
interaction strengths evolving
in accordance with the dynamical states of the elements.
We observe that in the presence of stochastic fluctuations in the
dynamics of the components, 
the system can exhibit large dispersion in the
time required for converging to the balanced state. This variability
is characterized by a
bimodal distribution, which points to an intriguing
non-trivial problem in the study of evolving energy landscapes.
\end{abstract}

\maketitle

\newpage
Many complex systems that arise in biological, social and technological
contexts can be represented as a collection of dynamical elements,
interacting via a non-trivial connection
topology~\cite{Newman10,Barrat08}. A variety of critical behavior has
been observed in such systems, both in the collective dynamics taking place
on the network, as well
as in the evolution of the network architecture
itself~\cite{Dorogovtsev08}.
The interplay between changes to the connection
topology (by adding, removing or rewiring links) and nodal dynamics
has also been investigated in different
contexts~\cite{Jain01,Gross06,MacArthur10,Durrett12,Goudarzi12,Liu12}.
While the coevolution of network structure and nodal activity
has mostly been studied in the simple case where the links are either
present or absent,
many naturally occurring networks have links with heterogeneously
distributed properties. Connections in such systems can differ
quantitatively by having a distribution of
weights (which may represent the strength of
interaction)~\cite{Barrat04,Onnela07}
and/or qualitatively through the nature of their interactions, viz.,
positive (cooperative or activating) and negative
(antagonistic or inhibitory)~\cite{Traag09}.
The presence of negative links in signed networks can introduce
frustration through the presence of inconsistent relations within
cycles in the system~\cite{Fischer91}. 
Networks whose positive and negative links are arranged such that
frustration is absent are said to be {\em structurally balanced} -- a
concept that was originally introduced in the context of social
interactions~\cite{Heider46}.
A classic result in graph theory is that 
a balanced network can be always represented as comprising two
subnetworks, with only positive interactions within each subnetwork,
while links between the two are exclusively
negative~\cite{Cartwright56}. Networks of dynamical elements 
with such structural organization can exhibit non-trivial collective
phenomena, e.g., ``chimera" order~\cite{Singh11}.

Recently, the processes through which structural balance can be achieved 
in networks has received attention from scientists and
quantitative models for understanding their underlying 
mechanisms have been proposed. 
Evolving networks where the sign of links are flipped to reduce
frustration have been shown to reach balance; however,
introduction of constraints can sometimes result in jammed states
which prevent convergence to the balanced
state~\cite{Antal05,Marvel09}.
Another approach, using coupled differential equations for describing
link adaptation~\cite{Kulakowski05}, has been analytically
demonstrated to result in balance~\cite{Marvel11}. 

While most studies on structural balance have been done
in the context of social networks, an important question is whether
other kinds of networks, in particular, those that occur in biology,
exhibit balance. The recent observation that the resting human brain
is organized into two subnetworks that are dynamically anti-correlated
(with the activity within each subnetwork being correlated)~\cite{Fox05} 
point to the intriguing possibility
that the underlying network may in fact be balanced.
As connections in the brain evolve according to long-term potentiation
which embodies Hebb's
principle~\cite{Hebb49}, i.e., the link weights change in proportion
to the correlation between activity of the connected elements,
it suggests a novel process for achieving structural balance.
Thus, signed and weighted networks can remove frustration by adjusting
the weights associated with the links in accordance with the 
dynamical states of their nodes.
Such a local adaptation process has an intuitive interpretation in
social systems, viz., agents that act alike have their ties strengthened,
while those behaving differently gradually develop antagonistic
relations. In fact, Hebb's rule may apply more broadly to a large
class of systems, for example, in gene regulation networks where it has been
suggested that co-expression of genes can lead to co-regulation over
evolutionary time-scales~\cite{Fernando09}. 

\begin{figure}
\begin{center}
\includegraphics[width=0.99\linewidth,clip]{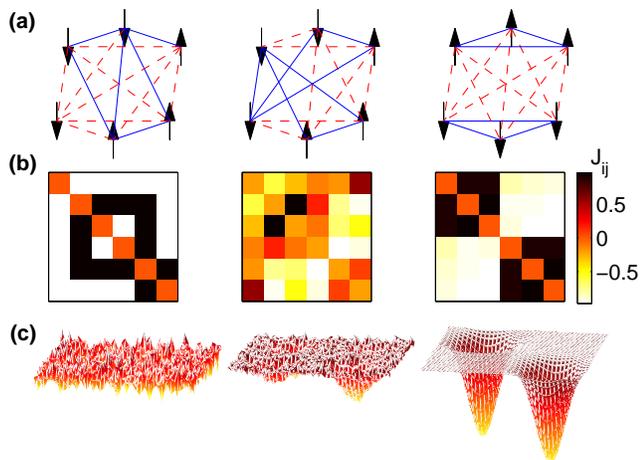}
\end{center}
\caption{Coevolution of coupling strength with the dynamics on the
node starting from a disordered state of spin orientations and
interaction strengths randomly selected to be $\pm 1$. 
(a) The spin configurations in the initial (left), intermediate
(center) and final, i.e., after convergence to structural balance
(right), states for a system of $N=6$ spins. Solid (broken) lines
represent positive (negative) interactions between spins.
The corresponding coupling matrices $J$ are shown in (b) while the
schematic energy landscapes are represented in (c). The two minima in
the balanced state correspond to the pair of degenerate ground states
related by reversal of each spin.
}
\label{fig1}
\end{figure}

In this paper, we show
that such a link-weight adaptation dynamics can in fact lead to
structural balance (shown schematically in Fig.~\ref{fig1}),
using only local information about the correlation between dynamical
states of the nodes. The temporal behavior of the approach
to balance shows unexpected features. In particular, we observe that
the system exhibits a high degree of variability in the time required
to converge to the balanced state when stochastic
fluctuations are present in the nodal dynamics. 
This relaxation time has a bimodal distribution 
for a range of adaptation rates and noise strengths.
Finite-size scaling of the transition from fast to slow
relaxation shows that the variation of the scaling exponent is related
to the qualitative nature of the way the bimodal distribution emerges.
As a larger
fraction of positive (negative) interactions reduces (promotes)
frustration, we also investigate the role of bias in the sign of
interactions on
the nature and rate of convergence to the balanced state.

We consider a system of $N$ globally coupled Ising spins $\sigma_i =
\pm 1$ ($i = 1, \ldots, N$), the energy for a given configuration of
spins being
\begin{equation}
\mathcal{E} = - \sum_{i \ne j} J_{ij} \sigma_i \sigma_j
\label{hamil}
\end{equation}
where $J_{ij} (= J_{ji}) $ is the symmetric bond, representing 
interaction strength 
between the spin pair ($i, j$).
Structural balance in real social networks have been recently investigated
using a similar energy function~\cite{Facchetti11}.
The balanced state corresponds to the situation where the interactions
are consistent with the corresponding spin pairs, i.e., $J_{ij}$ and
$\sigma_i \sigma_j$ have the same sign.
Starting from a disordered spin configuration and random distribution of
interactions, the state of the spins are updated stochastically at
discrete time-steps 
using the Metropolis Monte
Carlo (MC) algorithm with temperature $T$.
The interaction strengths also evolve after every MC step according
to the following deterministic adaptation dynamics:
\begin{equation}
J_{ij} (t+1) = (1-\epsilon) J_{ij} (t)  + \epsilon \sigma_i (t)
\sigma_j (t),
\label{Jijrule} 
\end{equation}
where $\epsilon$ governs the rate of change of the interaction
relative to the spin dynamics.
The $J_{ij}$ dynamics alters the energy landscape on which the state
of the spin system evolves. 
The {\em relaxation time} for the system is defined as the
characteristic time scale in which the balanced state is reached.
Note that the form of Eq.~(\ref{Jijrule}) ensures that 
the relaxation time $\sim 1/\epsilon$
in the absence of 
any thermal fluctuation (i.e., at $T = 0$).
Also, it restricts the
asymptotic distribution of $J_{ij}$ to the range $[-1,1]$, independent of
whether the system converges to a balanced state.
In many real systems the signature of the link cannot change,
although the magnitude of the link weight can.
We have also considered a variant of Eq.~(\ref{Jijrule}) 
for which the
dynamics is constrained such that the sign of each $J_{ij}$
cannot change from the initially chosen value. As a result 
several of the interactions can go to zero when the
system relaxes. 

In our simulations
the initial state of the
system 
for each realization
is constructed by choosing the spins $\sigma_i$ to be $\pm 1$
with equal probability. For most results shown here, each initial
$J_{ij}$ is chosen from a distribution with two equally weighted 
$\delta$ function peaks at $\pm 1$, i.e.,
$P (z; \mu) = [(1+\mu)/2] \delta (z-1) + [(1-\mu)/2] \delta (z+1)$
where the mean $\mu = 0$. We have verified that the results do not
change qualitatively if the initial distribution has a non-zero mean,
or has a different functional form (e.g.,
a uniform distribution in $[-1,1]$), 
provided that the system is initially far from balance.
For each set of parameters ($T, \epsilon$), 
$10^4$ different realizations have been
used to statistically quantify the relaxation behavior of the system,
which is identified using the energy per bond [Eq.(~\ref{hamil})]
normalized by the number of connections, i.e.,
$E =  \mathcal{E}/{N \choose 2}$,
as the order parameter.
The number of spins has
been chosen to be $N=64$ for most of the figures shown here, although
we have verified that the results are qualitatively unchanged for $N$ upto 512.
Simulating larger systems is computationally very expensive as the
system is globally coupled and disordered with time-varying
interactions.

In the absence of thermal fluctuations (i.e., at $T = 0$), the dynamics of
the system can be understood intuitively.
Starting from a random initial state, the spin dynamics stops
when the system gets trapped in local energy minimum within a few MC
steps ($\sim 1/\epsilon$, as mentioned above).
The subsequent evolution of the interaction strengths makes this
configuration a global minimum.
However, at finite temperature, the stochastic fluctuations of the
spins may prevent the system from remaining in a metastable state for
sufficiently long.
This does not allow the $J_{ij}$ dynamics to alter the energy
landscape sufficiently to make the configuration the global minimum.
Thus, an extremely long time may be required to reach structural
balance, and the relaxation time diverges due to the
stronger fluctuations on increasing temperature.

\begin{figure}
\begin{center}
\includegraphics[width=0.99\linewidth,clip]{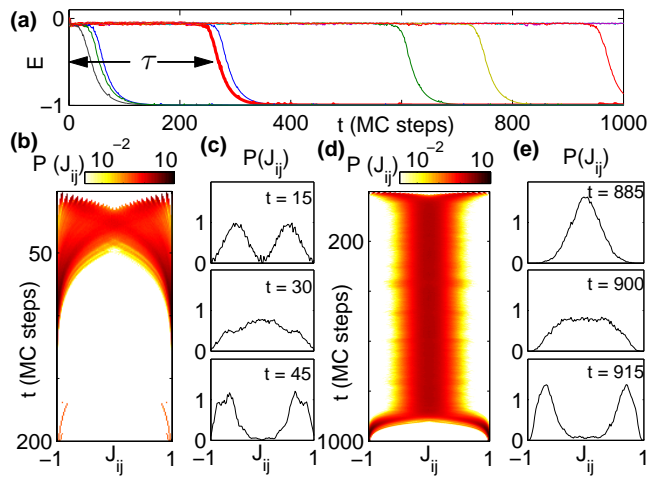}
\end{center}
\caption{(a) Typical time-evolution of the energy per bond $E$
for a system of $N$ spins
starting from different initial conditions.
The relaxation time $\tau$ indicated in the figure is the duration
after which $E$ decreases below $-0.5$. (b-e) Time-evolution 
of the distributions for the interaction strength $J_{ij}$ 
shown for two cases:
(b-c) when the system relaxes rapidly and (d-e) when convergence takes much
longer. Snapshots of the $J_{ij}$ distribution at specific times
immediately before, during and immediately after the convergence are shown 
for the two cases in (c,e) respectively. 
For all figures $N = 256$ with $T = 51$, $\epsilon = 0.05$.}
\label{fig2}
\end{figure}
Fig.~\ref{fig2}~(a) shows the time-evolution of the order parameter
$E$ for several typical runs for different initial conditions
and realizations of a system with $T = 15, \epsilon = 0.05$. 
The order parameter of the system initially corresponds to that for a maximally
disordered state ($\approx 0$) but eventually relaxes to a balanced
state ($E = -1$). The time required for reaching balance,
referred to as {\em relaxation time}, $\tau$, is
estimated by measuring the duration starting from the initial state
after which $E$ decreases below $-1/2$ [Fig.~\ref{fig2}~(a)].
For a large range of parameters, we observe two very
distinct types of
behavior: in one, the system relaxes rapidly, while in the other this
takes a longer time. 
In both cases, once the order parameter starts decreasing (i.e., after
time $\tau$), it reaches a balanced state within a time-interval 
$\sim 1/\epsilon$.
As this is
typically much shorter than the relaxation time for the second case,
the transition to the balanced state can appear rather suddenly
for the latter. Before the onset of the convergence to the balanced
state, the order parameter fluctuates over a very narrow range 
around zero, and there is little indication as to when the transition will
happen.
Characteristic time-evolution
corresponding to these two types of behavior are shown in 
Fig.~\ref{fig2}~(b-e). When the system relaxes rapidly, smaller peaks
emerge from the two peaks of the initial
$J_{ij}$ distribution (located at $\pm 1$) and eventually cross each
other to reach the opposite ends asymptotically, converging to a
two-peaked distribution again [Fig.~\ref{fig2}~(b-c)], indicating that all interactions are
now balanced. However, in the case where convergence takes
significantly longer [Fig.~\ref{fig2}~(d-e)], 
the initial distribution is first completely altered
to a form resembling a Gaussian distribution with zero mean. 
After a long time, the system abruptly converges towards a 
balanced state with a corresponding transformation of the $J_{ij}$
distribution to one having peaks at $\pm 1$.
Note that even with the same initial spin configuration and realization of
$J_{ij}$ distribution,
different MC runs generate distinct trajectories that
are similar to those shown in Fig.~\ref{fig2}~(a).
This implies that knowledge of the initial conditions is not
sufficient to
decide whether the system will relax rapidly or not.

\begin{figure}
\begin{center}
\includegraphics[width=0.99\linewidth,clip]{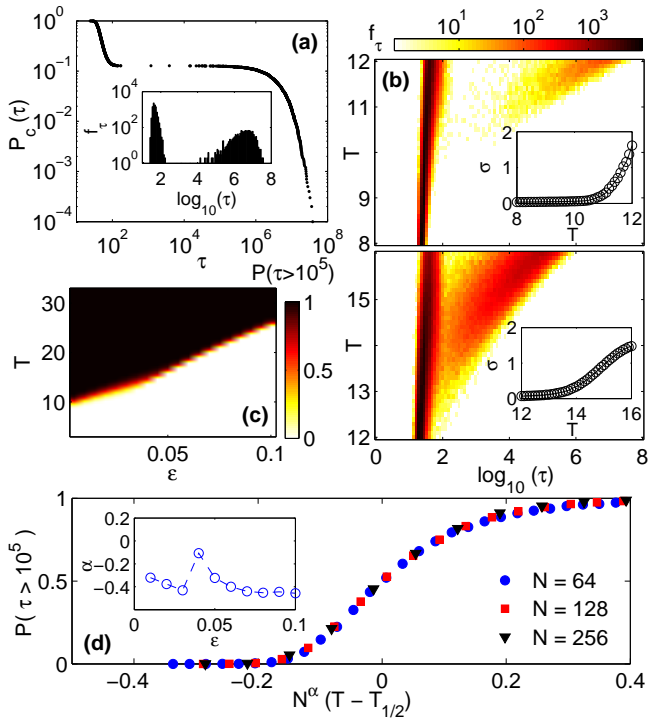}
\end{center}
\caption{(a) The cumulative distribution of relaxation time $\tau$ for
system of $N=64$ spins with $T = 12, \epsilon = 0.03$
shows a gap implying a bimodal nature for the distribution. The inset
showing the corresponding frequency distribution $f_{\tau}$ for $\log_{10}
(\tau)$ clearly indicates this bimodal nature.
(b) Probability distributions of $\log_{10} (\tau)$ shown as a function of
temperature $T$ for $\epsilon = 0.03$ (top) and 0.05 (bottom)
indicates the onset of bimodal
behavior at higher values of temperature, e.g., for $T \gtrsim 10$ in
(top). Bimodality appears around the temperature
where the standard deviation of $\log_{10} (\tau)$ starts
increasing appreciably from an almost constant value (insets).  
(c) The probability that relaxation takes longer than $10^5$ MC steps,
$P(\tau > 10^5)$ shown as a function of $\epsilon$ and $T$. The point
of transition from fast to slow convergence can be quantified by
$T_{1/2} (\epsilon)$, i.e., the temperature at which $P(\tau >
10^5)$=1/2 for a given value $\epsilon$ (indicated by boundary between
the dark and light regions).
(d) Finite size scaling of the probability that relaxation takes
longer than $10^5$ MC steps, $P(\tau > 10^5)$, with $N^{\alpha}
(T-T_{1/2})$ 
for different system sizes $N$ ($\epsilon = 0.05$). The temperature
at which $P(\tau > 10^5)$ becomes half is represented as $T_{1/2}$.
A scaling exponent value of $\alpha \approx -0.32$ shows 
reasonable data collapse.
The inset shows the scaling exponents for the best data collapse at 
different values of $\epsilon$.
%
}
\label{fig3}
\end{figure}
To quantitatively characterize the distinction between the two types 
of relaxation behavior, we focus on the statistics of $\tau$
(Fig.~\ref{fig3}). Fig.~\ref{fig3}~(a) shows the
distribution of the relaxation time for a given set of $(T, \epsilon)$ 
where cases
of both fast and slow convergences are seen. The resulting bimodal
nature is clearly observed with the peak at lower $\tau$ ($\sim 100$
MC steps) corresponding to fast convergence to balanced state while
that occurring at a higher value ($\sim 10^7$ MC steps) arises from
the instances of slow relaxation. The distribution decays
exponentially at very high values of $\tau$.
Fig.~\ref{fig3}~(b) shows the temperature dependence of the 
distribution of $\log_{10} (\tau)$ for two different values of
$\epsilon$. For the smaller $\epsilon$ (= 0.03), the second peak is
well-separated from the first when bimodality first appears, while for
the larger $\epsilon$ (= 0.05) the second peak appears close to the
first one.
To estimate the temperature where the second peak appears, we plot the
standard deviation of $\log_{10} (\tau)$ as a function of $T$ (inset),
as bimodality is characterized by an increase in the dispersion of
relaxation times.
To observe how the distribution is affected by variation in both $T$
and $\epsilon$, we show in Fig.~\ref{fig3}~(c) how the probability
that the relaxation takes a long time (viz., $\geq 10^5$ MC steps) varies as a
function of these two parameters. As we know that the system relaxes
rapidly when the temperature
is decreased close to zero, 
we expect this probability to be negligible at very low values of $T$.
On the other hand, when temperature is increased to very high values,
the relaxation takes increasingly longer, so that the probability
$P (\tau > 10^5)$ approaches 1. We indeed observe a monotonic increase
in this probability from 0 to 1 as the temperature is increased for a
given value of $\epsilon$. We can define a transition temperature
$T_{1/2} (\epsilon)$ as the value of $T$ at which this probability is 
equal to $1/2$. We observe that $T_{1/2} (\epsilon)$
increases with $\epsilon$, which implies that 
the relaxation to the balanced
state requires a longer duration as the
interaction dynamics becomes slower.
For a given $\epsilon$, we study the variation of the probability $P (\tau >
10^5)$ with $T$ for different system sizes.
Finite-size scaling shows data collapse with a scaling exponent
$\alpha$ [Fig.~\ref{fig3}~(d)] that varies with $\epsilon$ (inset).
Depending on the value of $\epsilon$, we observe that there may be
different types of bimodal distribution of the relaxation times, e.g.,
one where the second peak is clearly separated from the first, 
and the other where they are joined
[Fig.~\ref{fig3}~(b)]. The variation of $\alpha$ with $\epsilon$
appears to reflect this change from one type of bimodality to another
(inset).
\begin{figure}
\begin{center}
\includegraphics[width=0.99\linewidth,clip]{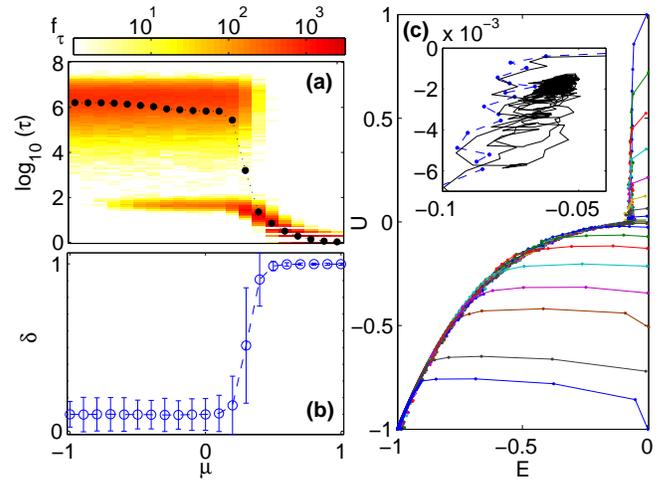}
\end{center}
\caption{
(a) Probability distribution of $\log_{10} (\tau)$ shown as a
function of the mean $\mu$ of the initial distribution for $J_{ij}$ for
$T = 17$, $\epsilon = 0.05$. The filled circles represent the average
of $\log_{10} (\tau)$ for different values of $\mu$.
The distribution does not change much for small bias ($|\mu|$); however
the lower peak disappears as $\mu$ approached $-1$ while the 
the relaxation behavior occurs faster as $\mu$ approaches $+1$. 
(b) Scaled size difference $\delta = (C_1 - C_2)/N$ between the two
clusters of aligned spins shown as a
function of $\mu$. As $\mu$ increases from negative values to 1,
$\delta$ increases from values close to 0 towards 1.
(c) Trajectories representing the time-evolution of the system
($N=256$) in the ($E,U$) order parameter space for different values of
$\mu$ (from top to bottom, $\mu$ increases from $-1$ to $1$ in steps
of $0.1$). After transients, all trajectories converge to a single
curve independent of the time required to converge to
the balanced state. 
A magnified view (inset) compares the trajectory 
corresponding to a long relaxation time (solid curve), which appears to be
trapped in this region, with the one corresponding to a short
relaxation time (broken curve)
for $\mu = 0$.
}
\label{fig4}
\end{figure}

So far we have assumed that the initial $J_{ij}$ distribution is
unbiased (i.e., $\mu = 0$). However, having a higher fraction of
interactions of a particular sign can have significant consequences
for both the structure of the final balanced state and the time
required to converge to it.
To investigate the role of this initial bias among the interaction
strengths,
we consider a distribution with two differently weighted $\delta$
function peaks at $\pm 1$ (i.e., $\mu \neq 0$).
Fig.~\ref{fig4}~(a) shows the distribution of the relaxation times as
$\mu$ is varied over the interval [$-1,1$] with the parameters $T$,
$\epsilon$ chosen such that 
there is
a clear bimodal nature of the relaxation time distribution
for the unbiased case ($\mu = 0$).
If all the interactions are
anti-ferromagnetic ($\mu = -1$), the system is extremely frustrated
and the relaxation to a balanced state may take a long
time, whereas in the case where the interactions are all ferromagnetic
($\mu = 1$), the system is balanced to begin with. Thus, with
increasing $\mu$, we expect the relaxation time to decrease, which is
indeed observed;
in addition, the peak at higher values of $\tau$ disappears as $\mu$
approaches 1. On the other hand, when $\mu$ approaches $-1$, the
peak corresponding to shorter relaxation times is no longer present. 
The two clusters that comprise the final balanced state can have very
different size distributions depending on the bias in the initial
distribution of $J_{ij}$. For the unbiased case, the two clusters are
approximately of the same size. We observe from Fig.~\ref{fig4}~(b)
that this property holds for the entire range of negative values for
$\mu$. As $\mu$ increases from 0, the size difference between the
two clusters start increasing, eventually leading to a single cluster
where all the spins interact with each other ferromagnetically ($\mu
\simeq 1$). Note that if the system initially has a very low degree of
frustration [e.g., $\mu \geq 0.4$ in Fig.~\ref{fig4}~(a,b)], the
system relaxes almost immediately to a balanced state where the larger
cluster comprises almost the entire system.
To visualize the coevolving dynamics in the link weights and spin
orientations as the system approaches balance for different values of
$\mu$,
we use an additional order parameter~\cite{Antal05,Marvel09} that measures the frustration in
a signed network in terms of the fraction of triads deviating from balance
(a triad being balanced if the product of its link weights approaches
$+1$),
$U = - \sum_{i,j,k} J_{ij} J_{jk} J_{ki} / {N \choose 3}$.
Fig.~\ref{fig4}~(c) shows that the trajectories corresponding to
different values of $\mu$ converge to a single curve after transients,
eventually reaching the balanced state at ($E=-1,U=-1$). For $\mu <
0$, the initial trajectory is approximately vertical indicating that
it is dominated by the adaptation dynamics (Eq.~\ref{Jijrule}), whereas
for $\mu > 0$, it has strong horizontal component implying that it is
governed primarily by the MC update of the spin states.
Realizations in which the system takes a long time to relax to the
balanced state
are distinguished by
trajectories that appear to be trapped in a confined region in the ($E,U$)
space for a considerable period [Fig.~\ref{fig4}~(c), inset].

%
%

We can qualitatively understand the appearance of short
relaxation times as follows.
In the initial state, when the system has a random assignment of
interaction strengths, the
energy landscape is extremely rugged, resembling that of a spin
glass~\cite{Fischer91}. The system starts out in a potential well
corresponding to one of the many initially available local minima.
As the state of the system evolves, the $J_{ij}$
dynamics (Eq.~\ref{Jijrule}) lowers the energy of the state by making
the interactions consistent with the spin orientations of the system,
while
the spin dynamics (updated according to the MC algorithm) can either
result in a further lowering of energy as the state moves towards the
bottom of the potential well, or is ejected from the initial local
minima due to thermal fluctuations. The probability of escaping from
the well at the $t$-th iteration, $p (t)$, depends on the potential
barrier height with neighboring wells. 
If the state cannot escape in the first few iterations 
from the local minimum from which 
it starts, successive lowering of the energy of this well by the 
$J_{ij}$ dynamics results in the minima becoming deeper, so that the
probability of escape is reduced further. Eventually, the system
relaxes to the balances state with a time-scale of $\sim
\epsilon^{-1}$, when the well
becomes the global minimum of a smooth energy landscape.
On the other hand, if the state escapes from the initial well within
the first few iterations, when the $J_{ij}$ dynamics has not yet been
able to significantly reduce the energy of a particular well, the
barrier heights separating the different local minima are all
relatively low. As a result, the system can jump from one well to
another with ease, corresponding to frequent switching of the
spin orientations. 
As $J_{ij}$ moves towards $\sigma_i \sigma_j$ at any given time
(Eq.~\ref{Jijrule}), rapid changes in the sign of the latter implies
that there is effectively no net movement of $J_{ij}$ towards $\pm 1$.
In fact, in this case, we observe 
that the initial
distribution of $J_{ij}$, comprising delta-function peaks at $\pm 1$, 
transforms within a few iterations to one resembling a Gaussian 
peaked at zero [Fig.~\ref{fig2}~(d-e)].
Once the system reaches such a state, it can only attain a balanced
state through a low-probability event which corresponds to the state
remaining in the same local minimum for several successive time steps.
As such an event will only happen after extremely long time, this will
lead to a very large relaxation time for a range
of $T$ and $\epsilon$.
Let us assume for simplicity that when the system is in the state
corresponding to frequent spin flips and low interaction strengths, the
probability of escaping from a local minimum is approximately a
constant ($p(t) \approx p$).
Then the probability that the system jumps between
different minima for $t$ steps and gets trapped in the $t+1$-th step is
$p^t (1-p)$. This results in the distribution of the relaxation
times (under the simplifying assumption of constant $p$) having an 
exponential tail, which is indeed observed [Fig.~\ref{fig3}~(a)].

%

To conclude, we have shown that a link adaptation dynamics inspired by
the Hebbian principle can result in an initially frustrated network achieving
structural balance. However, in the presence of fluctuations, 
we observe that
the system exhibits a large dispersion in the time-scale of relaxation
to the balanced state, characterized by a bimodal distribution. 
This extreme variability of the time required to converge to the
balanced state is a novel phenomenon that requires further
investigation. Our result suggests that even when a system has the
potential of attaining structural balance, the time required for this
process to converge may be so large that it will not be observed in
practice. 
Although we have considered a globally connected
network of binary state dynamical elements, it is possible to extend
our analysis to sparse networks~\cite{Dasgupta09} and different kinds of nodal dynamics
(e.g., $q$-state Potts model). As many networks seen in nature have
directed links, a generalization of the concept of balance to directed
networks and understanding how it can arise may provide important insights on the
evolution of such systems.

We thank Chandan Dasgupta, Deepak Dhar, S. S. Manna, Shakti N. Menon and
Purusattam Ray for helpful discussions. This work was supported in
part by CSIR, UGC-UPE, IMSc Associate Program and
IMSc Complex Systems Project.
We thank IMSc
for providing access to the ``Annapurna'' supercomputer.

\end{document}